\documentclass[12pt]{article}
\usepackage{times}
\usepackage{geometry}
\usepackage{natbib, graphicx, floatrow}
\usepackage[utf8]{inputenc}
\usepackage{enumitem,amssymb}
\usepackage{ragged2e}
\usepackage[export]{adjustbox}
\usepackage[font=footnotesize,labelfont=bf]{caption}
\newlist{thematic}{itemize}{8}
\setlist[thematic]{label=$\square$}
\usepackage{pifont}
\usepackage{tcolorbox}

\usepackage{aas_macros}

\usepackage{ulem}

\begin{document}
\raggedright
\huge
Astro2020 Science White Paper \linebreak

Gravity-wave asteroseismology of intermediate- and high-mass stars \linebreak
\normalsize

\noindent \textbf{Thematic Areas:} 
  $\boxtimes$  Stars and Stellar Evolution \linebreak
  
\textbf{Principal Author:}

Name: Andrew Tkachenko
 \linebreak						
Institution: Institute of Astronomy (IoA), KU Leuven, Celestijnenlaan 200D, 3001, Leuven (BE)
 \linebreak
Email: andrew.tkachenko@kuleuven.be
 \linebreak
Phone: +32-16-32.28.68 
 \linebreak
 
\textbf{Co-authors:} Conny Aerts, Dominic M. Bowman, Timothy Van Reeth, Joris De Ridder, Cole Campbell Johnston, May Gade Pedersen, Siemen Burssens, Mathias Michielsen, Joey Mombarg, Sanjay Sekaran, Robin Bj{\"o}rklund (all IoA, KU Leuven, BE), Tamara Rogers, Philipp Edelmann, Rathish Previn Ratnasingam (all Newcastle University, UK), Konstanze Zwintz (University of Innsbruck, AT), Juna Kollmeier (Carnegie Observatories, US), Jennifer Johnson (Ohio State University, US), Hans-Walter Rix (Max Planck Institute for Astronomy, DE), Jamie Tayar (University of Hawai'i, US)
  \linebreak

\begin{center}
\includegraphics[width=15cm]{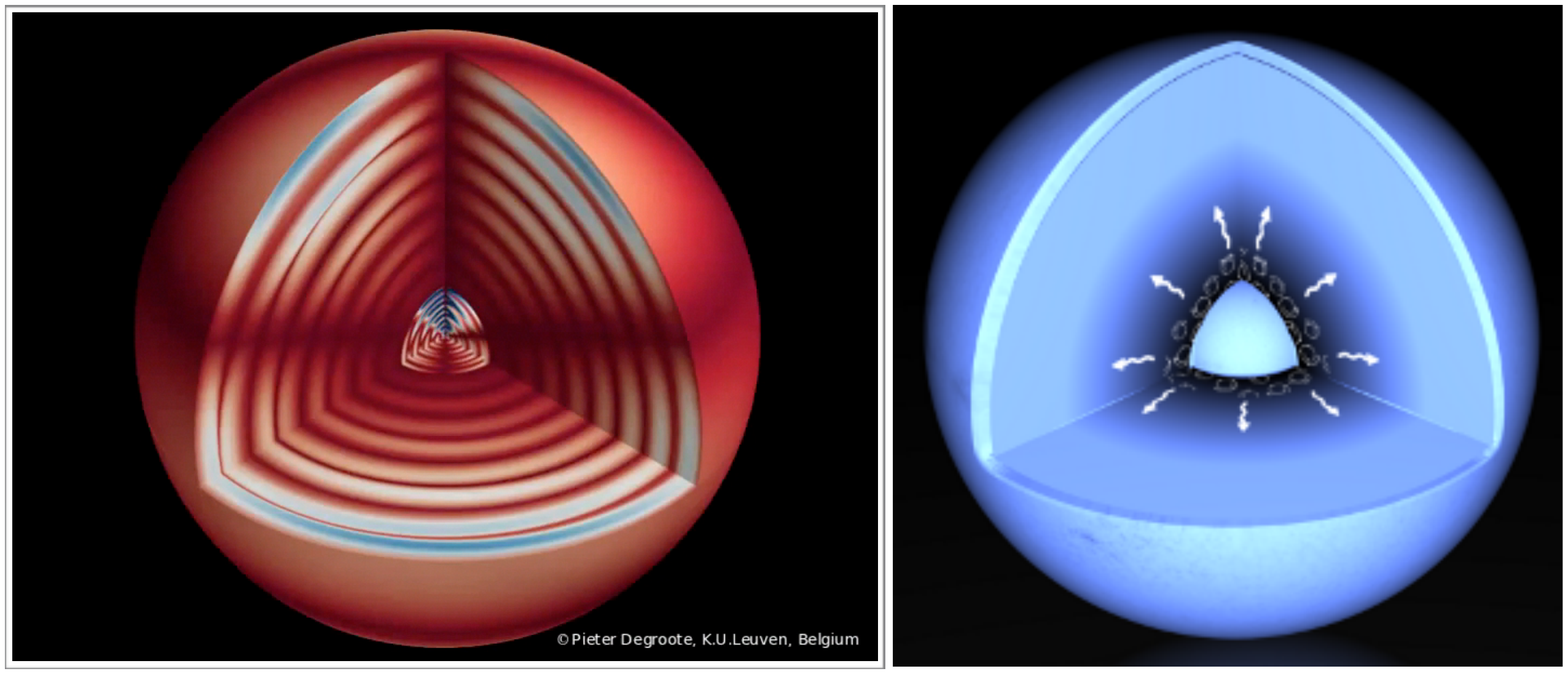}    
\end{center}

\pagebreak
\justify
\pagebreak

\section{Introduction}

Stars are the building blocks of galaxies, clusters, associations, and planetary systems. The properties that stars have at birth  define their complete evolutionary path until their deaths. The evolution of a star is driven by the physical processes in its interior making the theory of stellar structure and evolution (SSE) the most crucial ingredient for not only stellar evolution studies, but any field of astronomy which relies on the yields along stellar evolution. From the viewpoint of predicting the type of remnant that is left at the end of stellar life -- a white dwarf for stars with birth mass below $\sim$8~M$_{\odot}$ and a black hole/neutron star for stars with birth mass above $\sim$8~M$_{\odot}$ -- the theory of stellar evolution is well defined \citep[e.g.,][]{Kippenhahn2012}. However, high-precision time-series photometric data assembled by recent space missions revealed that current models of SSE show major shortcomings already in the two earliest nuclear burning phases, impacting all subsequent phases prior to the formation of the end-of-life remnant.

Inaccuracies in SSE models are largely due to the fact that their calibration has historically been restricted to matching stellar surface properties, such as effective temperature, surface gravity, surface abundances, and rotation. Although these ``classical'' observables depend on the processes active in the stellar interior, they do not provide direct ``in situ'' probing power of these processes. This limitation has changed dramatically with recent advances from {\it asteroseismology} -- the study of stellar interiors through the interpretation of the seismic variability of stars (see Sect~\ref{Sec: Asteroseismology}) -- largely driven by high-quality, high duty cycle, long-duration photometric data delivered by space missions. Stellar interiors can nowadays be studied with a level of detail that is inaccessible from surface measurements. The high-precision asteroseismic data reveal major shortcomings in SSE models, regardless of the stars’ evolutionary stages. {\it This white paper focuses specifically on the transport of chemical elements and of angular momentum (AM) in the SSE models of stars born with convective core, as revealed by their gravity-mode oscillations.} Following a brief introduction into asteroseismology as applied to gravity-mode pulsators and the concept of asteroseismic modeling, this paper highlights a few selected recent advances in the field. We ultimately outline future prospects for envisioned advances in observational, computational, and theoretical stellar astrophysics. For an extensive review of the problem of AM transport in stellar interiors the reader is referred to \citet{Aerts2019}.

\section{Asteroseismic modeling}
\label{Sec: Asteroseismology}

Asteroseismology, the interpretation of detected stellar oscillations \citep[e.g.,][]{Aerts2010}, offers a unique way to evaluate and calibrate stellar interiors. It uses observational evidence from the stellar interior rather than relying solely on measurements of stellar surface properties.

Following the linear theory of stellar oscillations, the displacement vector due to a non-radial oscillation mode of angular degree $l$ and of azimuthal order $m$ is written in terms of the spherical harmonic $Y^m_l$($\theta, \phi$) and has time dependence exp(--$i\omega t$), with $\theta$ the angle measured from the polar axis, $\phi$ the longitude, and $\omega$ the angular mode frequency. The radial component of four non-radial modes as seen by an observer under an inclination angle of 65$^{\circ}$ with respect to the rotation axis are illustrated below. 

\begin{center}
\includegraphics[width=8.5cm]{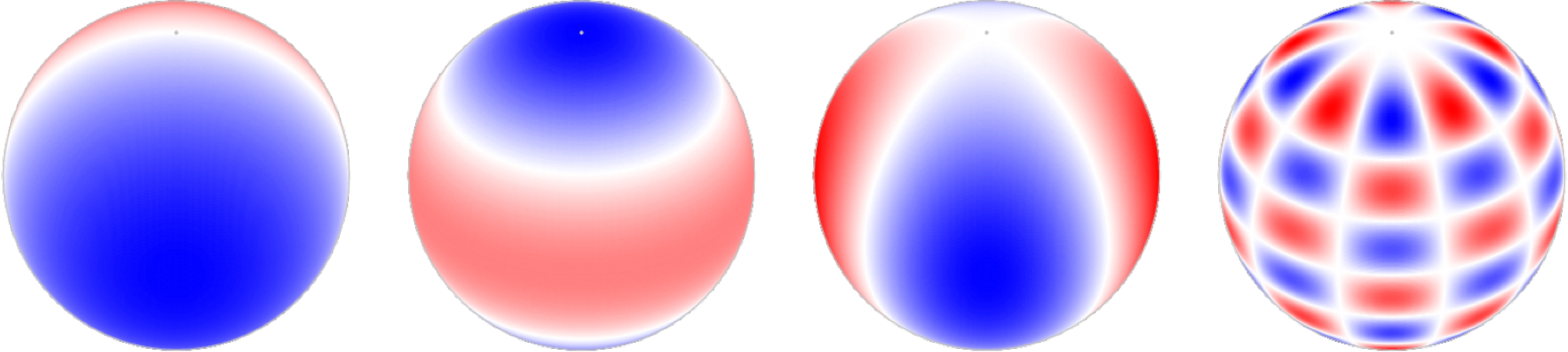}    
\end{center}

\noindent The white bands represent the positions of the surface nodes; red (blue) represent sections of the star that are moving in (out) at any given time in the oscillation cycle. From left to right: $(l, m)$ = 
(1,1), (2,0), (2,2), (10,5). The radial order $n$ of the mode represents the number of radial nodes of the radial componennt of the displacement vector in the stellar interior and cannot be deduced from surface observables. Rather, it must be estimated from comparison between observed and theoretically predicted frequencies with identified $(l, m)$, relying on stellar models. 

Stars that are born with a convective core (M\,$\gtrsim$\,1.4\,M$_{\odot}$) reveal {\it pressure (p-)} and/or {\it gravity (g-)}mode oscillations. In the case of the former, the pressure force acts as the dominant restoring force and the oscillations have  periods of a few hours. Such modes probe mostly the outer layers of stars. In contrast to p-modes, {\it g-modes are dominantly restored by buoyancy and have optimal probing power in the near-core regions of stars with a convective core.}

{\it Asteroseismic modeling} requires several observed and identified mode frequencies. These are fit with theoretically predicted frequencies deduced from stellar models. The latter rely on various assumptions for the microphysics of the stellar gas (chemical mixture, opacities, equation-of-state, etc.). The minimal free parameters to compute evolutionary models are the stellar mass $M$, the initial hydrogen and metal mass fractions $(X, Z)$, and the age $t$. Under conditions of moderate to rapid rotation, the effect of the Coriolis force must be taken into account in the computation of the g-mode pulsations. This requires knowledge of the interior rotation rate, $\Omega(r)$. Furthermore, interior chemical mixing in terms of convective overshooting, $D_{\rm ov}(r)$, occurs due to the inertia of convective elements, preventing them to stop abruptly at the convective boundary when entering a radiative layer. Hence, they ``overshoot'' into the radiative zone. In addition, chemical element mixing occurs in the radiative envelope, denoted here as $D_{\rm mix}(r)$, due to microscopic and macroscopic phenomena. Asteroseismic estimation of $\Omega(r)$, $D_{\rm ov}(r)$,  and $D_{\rm mix}(r)$ is done by assuming the simplest form of these three phenomena, i.e., for constant values of 
$\Omega$, $D_{\rm ov}$, and $D_{\rm mix}$ throughout the envelope. When this assumption turns out to be too simplistic to explain the measured mode frequencies, more complex forms of the rotation and mixing profiles
may be considered. 

The asteroseismic modeling involves statistical parameter estimation and stellar model selection. This is achieved from estimation of $M, t, X, Z, \Omega(r), D_{\rm ov}(r)$, and $D_{\rm mix}(r)$ and their uncertainties, by demanding the model prediction of these quantities to fit the observed values to within all available observational constraints, such as the  identified oscillation frequencies, spectroscopic effective temperature, astrometry-based radius and luminosity, surface abundances, etc. Bayesian stellar model selection is then performed with the aim to select the optimal input physics choices, keeping in mind the degrees of freedom in the estimation \citep[cf.,][for details]{Aerts2018}. 


\section{Asteroseismology of intermediate- and high-mass stars}

\begin{figure}
    \centering
    \includegraphics[width=16.5cm]{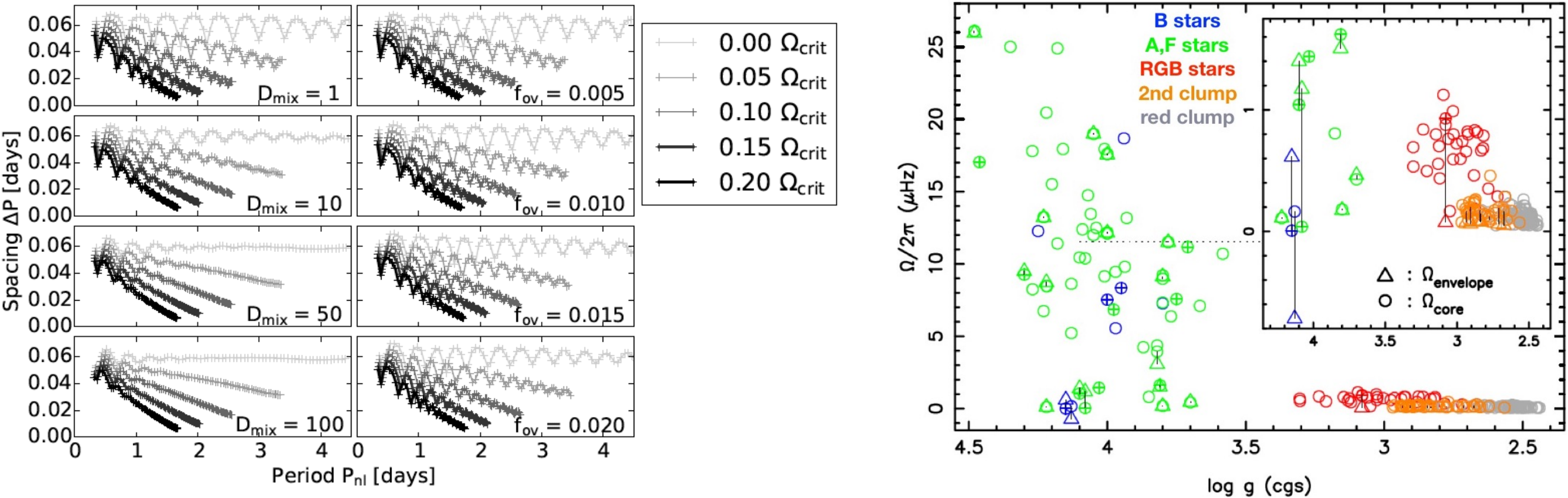}    
    \caption{{\bf Left:} Period spacings $\Delta P$ versus mode period $P_{nl}$ for prograde (i.e., traveling in the direction of stellar rotation) dipole ($l=1$) modes of consecutive radial order for a stellar model of 3\,M$_{\odot}$ and central
        hydrogen fraction $X_c=0.5$, for various constant rotation rates (expressed as fraction of the critical Roche rate), exponential overshooting with parameter values $f_{\rm ov}$ in the notation by \citet{Paxton2011}, and constant envelope mixing values $D_{\rm mix}$ (in cm$^2$s$^{-1}$). {\bf Right:} Core (circles) and surface (triangles) rotation rates (errors are smaller than the symbol size) as a function of surface gravity (error bar indicated for only one star as dotted line for clarity of the graph) for core-hydrogen burning stars with a mass between 1.4 and 2.0\,M$_{\odot}$ (green) and 3 to $\sim$5\,M$_{\odot}$ (blue), and for evolved stars of different evolutionary stages: RGB stars (red), red clump stars (grey) and secondary clump stars (orange). Binary components are indicated with an extra “+” sign inside the circles. Figure reproduced from \citet{Aerts2017}.}
    \label{fig:SpacingsRotation}
\end{figure}

High radial order g-modes are theoretically predicted to reveal characteristic period spacing patterns \citep[see, e.g.,][]{Miglio2008,Bouabid2013} whose properties are sensitive to the physical conditions in the near-core region of the star. This is illustrated in Figure~\ref{fig:SpacingsRotation} (left panel), which shows dipole g-mode period spacing patterns predicted for a model of 3~M$_{\odot}$ and central hydrogen fraction $X_c=0.5$, for various rotation rates and adopting the Traditional Approximation of rotation following \citet{Townsend2003}. The effects arising from various levels of exponentially decaying core overshooting (expressed by the parameter $f_{\rm ov}$) and different levels of constant diffusive envelope mixing $D_{\rm mix}$ \citep[both following][]{Paxton2011} are shown. The near-core rotation rate of the star sets the ``slope'' of the period spacing pattern, while its ``dips'' are caused by mode trapping due to the chemical gradient resulting from the shrinking core while $X_c$ decreased from the initial value of 0.7 to the adopted one of 0.5. This gradient alters the buoyancy frequency $N(r)$ outside the convective core and this
impacts upon the g-mode periods and their spacing pattern.

The first discovery of g-mode period spacings in core-hydrogen burning stars was achieved from a 150-day long CoRoT light curve \citep[][]{Degroote2010}. The precision of this spacing pattern was insufficient to deduce unambiguous mode identification (in terms of the degree $l$ and azimuthal order $m$), preventing detailed asteroseismic modeling. The latter was meanwhile achieved thanks to the ten times longer light curves observed by the Kepler mission \citep[][]{Kurtz2014,Papics2014,Papics2015,Saio2015,VanReeth2015,Keen2015,Murphy2016,Ouazzani2017,Papics2017,Szewczuk2018,GangLi2019}. Subsequent modeling of these intermediate-mass g-mode pulsators 
allowed estimation of near-core and core-to-surface rotation in F-type
\citep[][]{Tkachenko2013,Schmid2016,VanReeth2016,VanReeth2018} and B-type  \citep[][]{Triana2015,Moravveji2016} stars.
Figure~\ref{fig:SpacingsRotation} (right panel) provides a summary of this assessment (until 2017), where the core (circles) and envelope (triangles) rotation rates are depicted as a function of (spectroscopic) surface gravity (used as a proxy of stellar age) of the stars with mass in the range $1.4\,M_{\odot} < M < 5\,M_{\odot}$. A sample of evolved stars with masses below $\sim 3\,M_{\odot}$ is added for completeness. The measurements of both rotation rates and surface gravities for these evolved stars come from their detected mixed modes that have a p-mode character in the stellar envelope and a g-mode character in the deep interior of red giants \citep[][for reviews]{Chaplin2013,Hekker2017}. The character of mixed modes not only allows to identify the ongoing nuclear burning phase of red giants – something that cannot be done from surface measurements \citep[e.g.,][]{Bedding2011,Mosser2014} – but also how they rotate internally \citep[][]{Beck2012,Mosser2012,Deheuvels2014}. An update of the right panel of Figure~\ref{fig:SpacingsRotation} is available in Fig.\,4 of \citet{Aerts2019} and led to two major conclusions: 1) the cores of evolved stars rotate up to about ten times faster than their envelopes, which is  {\it about two orders of magnitude slower than standard theory of AM transport predicts}; 2)   intermediate-mass main-sequence stars rotating between a few \% and about 50\% of their critical Roche rate are near-rigid rotators. This implies  that {\it the cores of low- and intermediate-mass stars lose far more AM than predicted by the current theory of stellar evolution.} 

\citet{Aerts2019} provide an overview of currently known AM transport mechanisms. Following up on this, \citet{Fuller2019} put forward a better version of the magnetic Tayler instability. The asteroseismic signature of the interior magnetic field connected with this instability (their Fig.\,8) on g-modes has so far not yet been evaluated, but work is ongoing to achieve this (Prat et al., in preparation). An alternative explanation of efficient AM transport is the one induced by Internal gravity waves (IGWs). These  propagate in regions that have a sub-adiabatic temperature stratification, i.e.\ in radiative regions. For the stellar mass range considered in this white paper, IGWs are excited stochastically at the interface of the convective core and the radiative envelope. Their propagation and dissipation in such stars were studied numerically by \citet{Rogers2013} and Edelmann et al. (submitted). IGWs have been detected in the form of a spectrum of low-frequency waves from CoRoT space photometry of three O-type stars by \citet{Aerts2015}, who also  demonstrated the effect of these waves on diagnostic atmospheric spectral lines of high-mass stars. These findings agree qualitatively with the detection of large macro-turbulent surface velocity fields in OB-type stars \citep{Simon-Diaz2014,Simon-Diaz2017}. This re-enforces the connection between IGWs and
macroturbulence observed for these stars, either through g-modes \citep[which are the standing gravity waves among the entire spectrum of IGWs][]{Aerts2014} or from the collective effect of the travelling IGWs, or both. Recently, \citet{Bowman2019a} and Bowman et al.\ (submitted) presented compelling evidence that about 75\% of more than hundred OB-type variables, including many blue supergiants, reveal prominent low-frequency stochastic variability in CoRoT, K2, and TESS space photometry. The detected signal cannot be explained in terms of surface granulation, while it is consistent with numerical predictions for IGWs. Bowman et al.\ (submitted) present IGW detections in tens of stars in the Large Magellanic Cloud (LMC) from TESS photometry. The underlying physical mechanism responsible for the detected low-frequency variability is independent of metalicity, which leads to the conclusion that it cannot be caused by 
sub-surface convection or heat-driven g-mode oscillations requiring a high metal opacity in the stellar envelope. IGWs excited at the convective core boundary offer a natural explanation for the detected low-frequency variability in space photometry and for the missing AM transport, suggesting it to be a common astrophysical phenomenon for these stars.

The co-existence of coherent g-modes and a spectrum of travelling IGWs provides optimal probing power in terms of interior mixing as a function of depth inside the star. As demonstrated by \citet{Pedersen2018}, coherent g-modes offer us the possibility to measure the amount and shape of core overshooting depending on the evolutionary stage. At the same time, the element transport due to IGWs \citep{Rogers2017} dictates the form of mixing towards the surface as a function between $[\rho(r)]^{-1}$ and $[\rho(r)]^{-1/2}$.

\section{The future of gravity-wave asteroseismology is bright}

The study of stellar interiors for intermediate- and high-mass stars has a bright future. We identify the key observational, computational, and theoretical advances necessary to push gravity-wave asteroseismology to the next level and to meet the ultimate goal of {\it improving the theory of AM and element transport}.

The pre-requisite for precision gravity-wave asteroseismology is to detect and identifying many individual frequencies of the standing gravity waves within the entire spectrum of IGW, irrespective of their excitation. This sets strict requirements on {\it photometric data}, namely high duty cycle, high signal-to-noise level, and sufficiently long time-base of 200 days or longer. These requirements are being satisfied the coming decade thanks to the ongoing and planned space missions TESS \citep{Ricker2016} and PLATO \citep{Rauer2014}. Optimal exploitation of asteroseismology, even for single objects, requires additional observational constraints in terms of the fundamental atmospheric properties, such as effective temperature, $v\sin\,i$, luminosity, radius, and atmospheric abundances, most notably for C, N, and O. These observational requirements have to be assembled for large representative samples of thousands of stars, implying the need of {\it all-sky, multi-epoch high-precision spectroscopy} in support of space photometry. Multiple-epoch spectroscopy offers the opportunity to link asteroseismic signatures of IGWs with time-dependent spectral-line broadening beyond $v\sin\,i$ and to exploit synergies between asteroseismology and binary stars. Binary asteroseismology allows to take advantage of two stars orbiting around their common centre of mass to extract their fundamental parameters, such as masses and radii, in a {\it model-independent\/} way and with relative precisions of $\sim$1\% \citep[see, e.g.,][for a modern methodological framework]{Prsa2016}. This opportunity occurs for detached, eclipsing, double-lined spectroscopic binaries, allowing for so-called {\it dynamical asteroseismology}. In addition, close binary systems, where the gravitational interaction between the two stars gives rise to tides that are sufficiently strong to alter each other's interior structure and hence impact their pulsation properties, offer the application of so-called {\it tidal asteroseismology}. By scaling the outcome of tidal asteroseismology down into the sub-stellar mass regime, opportunities arise to do asteroseismically-driven studies of exoplanets around intermediate-mass stars and their interactions with their host star. Given that many intermediate- and -high-mass stars are found in binary systems, with the binary fraction increasing rapidly towards higher masses and reaching $\sim$80\% for the most massive O-stars \citep{Sana2012}, {\it synergies between asteroseismology and binarity should be among the highest priority topics in stellar astrophysics for the next decade.}

The requirement of analysing large samples of stars in an efficient way, calls for robust methodologies to  select appropriate stellar samples in the first place, and for subsequent fast yet thorough modeling and interpretation of the results. Various machine learning-based approaches are currently being developed and exploited (e.g., \citealp{Armstrong2016}; \citealp{Valenzuela2018} for stellar variability classification; \citealp{Hendriks2019} for seismic modeling), setting the stage for future applications. Coupled with the necessity of computationally expensive 3D (magneto)hydrodynamical simulations of convection and IGWs and their subsequent implementation into 1D SSE models, the envisioned {\it advances in computational astrophysics connected with high-performance computing and data storage infrastructures} are an essential element for the development of gravity-wave asteroseismology as a prominent research field.

\pagebreak
\textbf{Acknowledgements}

AT, CA, JDR, DB, TVR, CJ, MP, SB, MM, and JM receive funding from the European Research Council (ERC) under the European Union’s Horizon 2020 research and innovation programme (grant agreement N$^\circ$670519: MAMSIE) and from the KU\,Leuven Research Council (grant C16/18/005: PARADISE), as well as from the Belgian Science Policy office Belspo: PLATO. 


\end{document}